\documentclass[twocolumn,floatfix,superscriptaddress,amsmath,showpacs,showkeys,aps,prb]{revtex4-1}
\usepackage[utf8]{inputenc}
\usepackage[final]{graphicx}
\usepackage{bm}
\usepackage[normalem]{ulem}
\usepackage[usenames]{color}
\usepackage{times}
\usepackage{amsmath}
\usepackage{verbatim}
\usepackage{amssymb}
\usepackage{xcolor}
\usepackage{hyperref}
\usepackage{physics}
\usepackage{wrapfig}
\usepackage{subfigure}
\DeclareGraphicsExtensions{.eps}

\begin{document}

\title{Probing Fermi liquid exceptional points through AC conductivity}
\author{Rui Aquino}
\affiliation{Departamento de F{\'\i}sica Te\'orica,
Universidade do Estado do Rio de Janeiro, Rua S\~ao Francisco Xavier 524, 20550-013  
Rio de Janeiro, RJ,  Brazil}
\author{Daniel G. Barci}
\affiliation{Departamento de F{\'\i}sica Te\'orica,
Universidade do Estado do Rio de Janeiro, Rua S\~ao Francisco Xavier 524, 20550-013  
Rio de Janeiro, RJ, Brazil}
\date{\today}

\begin{abstract}
Exceptional points, which are topological non-Hermitian degeneracies, show up in the collective mode spectrum of Fermi Liquids with high angular momentum interactions. In this paper,  we look for signatures of these non-trivial singularities by computing the AC conductivity of Fermi liquids with dipolar and quadrupolar interactions in a narrow slab. We show that the finite size of the slab imprints clear signatures of the structures of the collective mode spectrum in the conductivity as well as in the dephasing between the electric field and the current density in a wide range of coupling constants. 
In particular,  we show the fingerprints of exceptional points,  observed in the weak attractive dipolar and/or quadrupolar regime. 
The main result does not depend on the specific model of quasi-particle interactions.  We also discuss some actual compounds where these phenomena could be experimentally observed.

\end{abstract}

\maketitle

\section{Introduction}
\label{Sec:Introduction}
The Landau Theory of Fermi Liquids states the paradigms to describe the normal state of metals since the late 50's\cite{nozieres-1999, baym1991}. Although in the last 30 years states of matter which don't follow this paradigm has been in focus, Fermi Liquids still attracts interest and hosts unexpected phenomena. One of the most recently reported is the emergence of non-hermitian degenerancies, the so called Exceptional Points (EP), in the longitudinal polarization channel of the collective mode spectrum of a two-dimensional Fermi Liquid with quadrupolar interactions\cite{Aquino-2019, Aquino-2020}. 

Exceptional points are singularities of the Hilbert space characterized by  a level degeneracy where not only the energy eigenvalues coalesce but also the eigenvectors. Therefore, at these special points of the parameter space, the Hamiltonian cannot be diagonalized. 
This type of singularities cannot occur in Hermitian systems where, even at a degeneracy, the Hamiltonian eigenvectors should be orthogonal.  In correlated electron systems,  non-Hermitian phenomena can emerge in different ways. One example is the inelastic electron-electron or  electron-phonon scattering. These channels trigger damped collective excitations as well as EPs explicitly\cite{kozii2017}. Another example is the collective mode spectrum of Landau Fermi liquids.  
Collective modes are not a closed system, since they can exchange energy with individuals quasi-particles. In this way, damped modes appear in the spectrum. This mechanism, known as  ``Landau damping",  is the main source of non-Hermiticity.  

From the experimental perspective, the study of collective modes of strongly correlated systems is quite involved\cite{CastroNeto2020}. 
In principle, momentum-resolved dynamic susceptibility in the meV scale~\cite{Abbamonte2017} should give us information about collective excitations.  Another possibility could be the use of  pump-probe spectroscopy~\cite{Giannetti2016,Misha2009,Mitrano2019} to observe the dynamic response in the time domain.  Also, spatially nonlocal eletromagnetic response at low temperatures, such as the anomalous skin effect\cite{Reuter1948, dressel2002}, should encode these excitations as well. On the other hand, recent advances in the quality of crystal growth have allowed experimentalist to reach new regimes of metallic transport\cite{Bandurin2016, Moll2016, Gusev2018}. We will argue that these ultraclean samples are good candidates for obtaining  signatures of collectives modes. In particular, recent studies of microwave photoresistivity in ultraclean GaAs quantum wells\cite{West2010, West2011} reported singular resonances. The cause of these effects have been pointed out to be caused by transverse zero sound waves in highly nonideal Fermi Liquids, in the presence of a magnetic field\cite{Alekseev-2019, Alekseeva-2019}. This is the type of effect that we are looking for. 

In this article we look for signatures of the collective mode structure of a Fermi liquid and, in particular, signatures for EPs in transport properties.  Specifically, we compute the AC conductivity of a narrow slab. In principle, the bulk conductivity should not give information about collective modes since this observable is  essentially related with the  ``center of mass" dynamics. Thus, electron-electron forward scattering interactions could not play a relevant role. However, the finite width of the slab, $W$, introduces a natural scale of momentum $q_0\sim 1/W$ and, correspondingly, an energy scale $\omega_0=v_F q_0 $, where $v_F$ is the Fermi velocity. We expect that  the AC conductivity  could in principle provide signatures of collective modes provided their frequency $\omega\sim \omega_0$. Indeed, this kind of set up was recently used to study shear modes in Fermi Liquids with dipolar interactions\cite{Sodemann-2020}. In this work, we are interested in probing longitudinal polarized modes in a model with dipolar and quadrupolar interactions. In particular, a rich structure of collective modes was recently reported\cite{Aquino-2019, Aquino-2020} for quadrupolar modes. For dipolar interactions, we will show a similar structure in the same polarization channel of the collective modes.

Using the set up shown in Fig.  (\ref{fig:sample}), we have computed the frequency dependent current density using linear response in the context of Landau theory of Fermi liquids within a model of dipole and quadrupolar interactions.  The first result of the paper is given by Eq. (\ref{eq:sigmaqx0}), where we show an explicit expression for 
the AC conductivity $\sigma(\omega)$ in terms of the Green function of the longitudinal polarized dipolar modes, $G^+_{11}(\omega,{\bf q})$. We have exactly computed this Green function and 
we have numerically evaluated  the modulus of the conductivity $|\sigma(\omega)|$ and the dephasing angle $\varphi=\arctan[\Im(\sigma)/Re(\sigma)]$ for different values of the parameters of the model. We have analyzed three distinct regimes:  a strong repulsive regime in which there are only one stable collective mode,  the so called zero sound excitation (Fig. (\ref{fig:sigma1P})), a  weak attractive regime, in which there are two stable collective modes  (Fig. (\ref{fig:sigma2P})) and finally,  as our main result, we display in Fig.  (\ref{fig:sigmaEP}) signatures of the exceptional point,  where both stable modes coalesce in a double pole structure. 

We have confirmed that, depending on the boundary conditions on the walls of the slab, the conductivity and specially the dephasing angle is very sensitive to the presence of collective modes. The repulsive and attractive regions are clearly differentiated by the peak structure of the conductivity and the structure of the dips of the dephasing angle. 
Moreover,  the exceptional point imprints clear signatures in the complex AC conductivity. 

The paper is organized as follows. In Sec.~\ref{Sec:Transport} we define our set up and briefly review how to deal with transport in Fermi Liquid Theory. In Sec.~\ref{Sec:FL} we show how to introduce bulk and boundary conditions in the model and how to compute the current density.   We define our model of interactions and compute the conductivity as a function of Green functions in Sec.~\ref{Sec:Model}.  In Sec.~\ref{Sec:Collective} we compute the collective mode spectrum from the poles of the Green function and in Sec.~\ref{Sec:Conduct} we numerically compute the conductivity for different values of the parameters of the model. 
Finally we discussed our results in Sec.~\ref{Sec:discussions}.

\section{Transport in a narrow metallic slab}
\label{Sec:Transport}
Transport properties of Fermi liquids have been studied for a long time by using using the Landau-Silin-Boltzmann-equation approach\cite{Ashcroft-1982}. However, two-dimensional Fermi liquids with high angular momentum interactions have unique properties, specially related with the collective excitations.

\begin{figure}
	\begin{center}
		\includegraphics[width=0.5\textwidth]{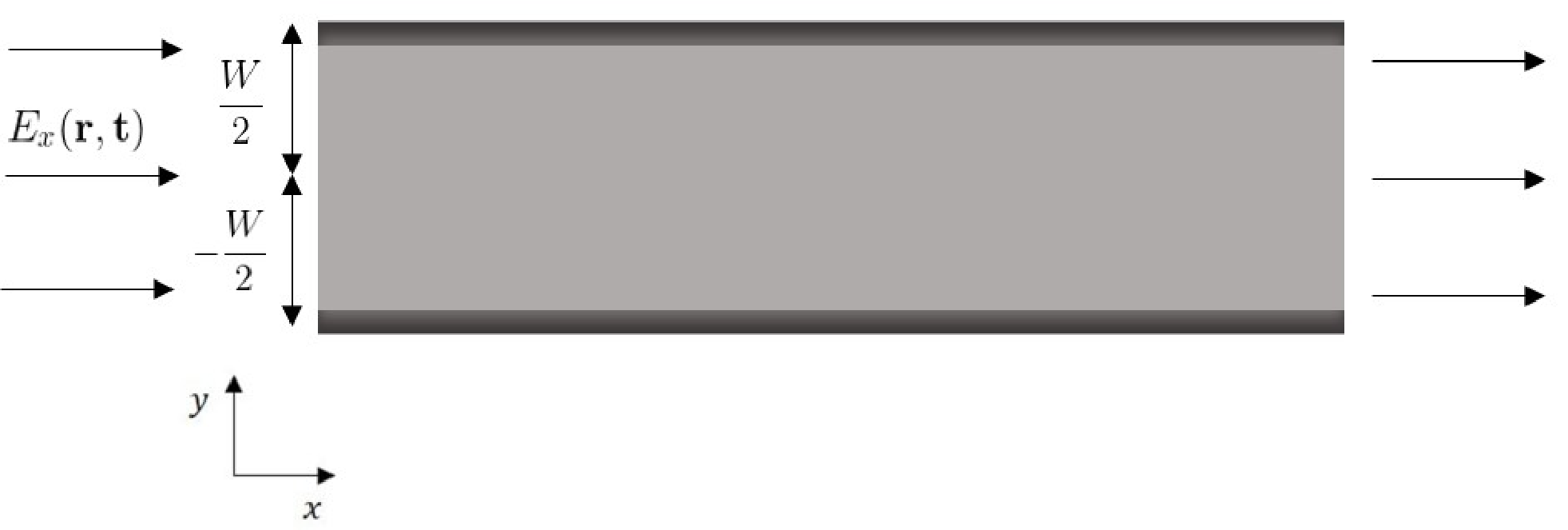}
	\end{center}
	\caption{ Two-dimensional Fermi liquid confined in a slab set up.  The width of the slab is $W$ and the electric field is applied in the horizontal direction.  The darker areas in the sample are the walls that we will describe  through  the collision integral $I_{\rm bd}$ in the transport equation}
	\label{fig:sample}
\end{figure}

We will focus on charge transport in a metallic slab in which we suppose that electron-electron interactions are well described by 
a Fermi liquid.  To be concrete, we consider a rectangular system,  as shown in Fig. (\ref{fig:sample}). 
Linear response provides the general form of the current density, 
\begin{equation}
J_i({\bf r},t)= \int d^2r'dt'  \sigma_{ij}({\bf r}, {\bf r}', t, t') E_j({\bf r}', t')
\label{eq:LinearResponse}
\end{equation} 
where $J_i$ is the $i^{\rm th}$ component of the current density, with $i= x,y$.  $E_j({\bf r}, t)$ is the $j^{\rm th}$ component of an applied electric field and the conductivity is given by the tensor $\sigma_{ij}({\bf r}, {\bf r}', t, t')$.
 
The system is  invariant under translations in the $x$ direction. On the other hand, it breaks this symmetry in the $y$ direction due to the boundaries of the slab. In this situation, and in the stationary regime, we expect that 
\begin{equation}
\sigma_{ij}({\bf r}, {\bf r}', t, t')\equiv \sigma_{ij}(x-x', y,y', t-t').
\label{eq:cond-tensor}
\end{equation}
Fourier transforming Eq. (\ref{eq:LinearResponse}),  we have
\begin{equation}
J_i(q_x, q_y, \omega)= \int dq_y ' \; \sigma_{ij}(q_x, q_y, q_y', \omega) \;E_j(q_x, q_y',  \omega).
\label{eq:Jslab}
\end{equation} 
Thus,  the presence of boundary conditions induces a non-local expression in  momentum space. 
In the absence of magnetic fields or magnetic impurities,  the conductivity tensor is diagonal.  Choosing the electric field in the $x$ direction (as shown in Fig. (\ref{fig:sample})), we are led with the computation of the non-local conductivity
\begin{equation}
\sigma({\bf q}, q_y', \omega)\equiv \sigma_{xx}(q_x, q_y, q_y', \omega)
\label{eq:conductivity}
\end{equation}
In addition,  if we consider that the contacts are uniformly distributed along the whole transverse section $W$,  then the electric field has the form $E_x({\bf q})=E_x(q_x) \delta(q_y)$.  In this situation the conductivity is a  local function of the momentum  $\sigma(\bf q,\omega)$.

In order to compute  the current density,  we take advantage of the usual Landau theory of Fermi liquids\cite{nozieres-1999, baym1991}.  Within this theory,  we can write the non-homogeneous local quasi-particle energy as
\begin{equation}
	\varepsilon_{\bf k}({\bf r},t) = \epsilon_{\bf k} + \sum_{\bf k'}f_{\bf k,k'}\,n_{\bf k'}({\bf r},t) - e\,\phi({\bf r},t)
\end{equation}
where $\epsilon_{\bf k}$ is the bare dispersion relation, $e$ is the quasi-particle charge, $n_{\bf k}({\bf r},t)$ is the non-homogeneous occupation number, $\phi({\bf r},t)$ is an external scalar potential and $f_{\bf k,k'}$ codifies short-range forward scattering  interactions (notice that this expression could describe a wider range of interactions in the almost ideal gas\cite{LifshitzPitaevskii-Book}). 
In the spirit of semiclassical description, the momentum $|{\bf k}|\sim k_{\sf F}$, where $k_{\sf F}$ is the Fermi momentum,  and the distance $r\gg 1/k_{\sf F}$. For simplicity, only the quasiparticles charge degree of freedom is taken in account. That means that  we can average over the spin variables, ignoring exchange and spin-orbit interactions.

The particle-hole dynamics is given by the Boltzmann equation  
\begin{align}
	&\frac{\partial n_{\bf k}({\bf r},t)}{\partial t} = \label{eq:Boltzman}  \\
	&\frac{\partial \varepsilon_{\bf k}({\bf r},t)}{\partial {\bf r}}\cdot\frac{\partial n_{\bf k}({\bf r},t)}{\partial {\bf k}} - \frac{\partial \varepsilon_{\bf k}({\bf r},t)}{\partial {\bf k}}\cdot\frac{\partial n_{\bf k}({\bf r},t)}{\partial {\bf r}} + I_{\rm coll}[n_{\bf k}],
\nonumber
\end{align}
where $I_{\rm coll}$ is the collision integral.   This is  a non-linear equation for $n_{\bf k}$.  
Within linear response, we can write $n_{\bf k}({\bf r},t) = n^0_{\bf k} + \delta n_{\bf k}({\bf r},t)$, where $n^0_{\bf k}=\Theta(\epsilon(\bf k)-\mu)$, with $\mu$ the chemical potential,  is the occupation number of  free fermions  at zero temperature and  $\delta n_{\bf k}({\bf r},t)$ are small perturbations strongly peaked  at $k\sim k_{\sf F}$.  Linearizing  Eq. (\ref{eq:Boltzman}) in $\delta n_{\bf k}$ and  
using  the definition of the Fermi velocity ${\bf v}_{\sf F}=\grad_{\bf k} \varepsilon_{\bf k}$, we find the reduced Boltzmann equation,
\begin{align}
	\frac{\partial \, \delta n_{\bf k}({\bf r},t)}{\partial t} &+ {\bf v}_{\sf F} \cdot \nabla \left[ \delta n_{\bf k}({\bf r}) -\frac{\partial n^0}{\partial \varepsilon_{\bf k}} \sum_{\bf k'}f_{\bf k,k'} \delta n_{\bf k'}({\bf r}) \right] \nonumber \\ 
	&- e\frac{\partial n^0}{\partial \varepsilon_{\bf k}}{\bf v}_{\sf F} \cdot{\bf E}({\bf r},t) = I_{\rm bk}[ \delta n] + I_{\rm bd}[ \delta n],\label{eq:bolt}
\end{align}
where $\partial n^0/\partial \varepsilon_{\bf k} = -\delta(\epsilon_{\bf k} - \mu)$. The delta function projects the momentum ${\bf k}$ into the Fermi surface ${\bf k=k}_{\sf F}$.  We assume a circular Fermi surface and,  since $r \gg 1/k_{\sf F}$,  boundary conditions do not significantly modify  this symmetric shape.  
There are two collision operators in the r.h.s of Eq.~(\ref{eq:bolt}); one of them describes electron-impurity scattering in the bulk ($I_{\rm bk}$),  and the other one describes collisions within the boundary of the sample ($I_{\rm bd}$). 

It is worth noticing that in our model we have not considered any interaction that could lead to a superconducting instability\cite{LifshitzPitaevskii-Book}. The competition between interactions in the  forward scattering particle-hole channel (described by Landau parameters) and in the particle-particle channel that could lead to superconductivity is an interesting subject\cite{Legget-1965}. We will not address this point in this paper since we are essentially interest in the metallic phase.

Once Eq. (\ref{eq:bolt}) is solved for $\delta n_{\bf k}({\bf r} ,t)$, the current density can be easily computed from 
\begin{equation}
{\bf J}({\bf r}, t)= e \sum_k {\bf  j}_{\sf F}\, \delta n_{\bf k}({\bf r},t)
\label{eq:J}
\end{equation}
where
\begin{equation}
{\bf j}_{\sf F} = {\bf v}_{\sf F} - \sum_{\bf k'}\frac{\partial n^0}{\partial \epsilon_{\bf k'}}\, f_{\bf k,k'}\, {\bf v}_{\sf F}'~.
\end{equation}

By explicitly computing  Eq. (\ref{eq:J}) and by comparing it  with Eq. (\ref{eq:Jslab}), we can read the AC conductivity. 

In the following  sections we  show how to work out expressions Eq. (\ref{eq:bolt}) and (\ref{eq:J}) in order to give them and explicit computational form.

\section{Fermi surface fluctuations,  collision integrals and the  angular momentum basis }
\label{Sec:FL}
It is convenient to parametrize fluctuations as,
\begin{align}
	\delta n_{\bf k}({\bf r},t) = \delta(\epsilon_{\bf k}-\mu)\,v_{\sf F}\,m_{\bf k}({\bf r},t)
	\label{eq:normal}
\end{align}
where $m_{\bf k}({\bf r},t)$ represent the normal displacement of the Fermi Surface at  the point ${\bf k=k}_F$. 
In order to Fourier transform from $m_{\bf k}({\bf r},t)$ to $m_{\bf k}({\bf q},t)$,  it is necessary to take into account the boundary conditions.   Assuming that the  electric field in the $x$ direction has  reflection symmetry $E_x(x,y)=E_x(x,-y)$,   it is simple to verify that $m_k(x, W/2, t)=m_k(x,-W/2,t)$. Thus,   the momentum 
is discretized in  the $y$ components,  $q_y=n q_0$, with $n=0,\pm 1, \pm 2, \ldots$ and $q_0=2\pi/W$. 
The Fourier transformed version of Eq.~(\ref{eq:bolt}) is, 
	\begin{align}
	&\frac{\partial\, m_{\bf k}({\bf q})}{\partial t} +i\, {\bf v}_{\sf F}\cdot{\bf q}\left[m_{\bf k}({\bf q})+\sum_{\bf k'}F_{\bf k,k'}\,m_{\bf k'}({\bf q}) \right]
\nonumber\\
	&=\frac{e}{v_{\sf F}}\,{\bf v}_{\sf F}\cdot{\bf E}({\bf q}) + I_{\rm bk}[m_{\bf k}({\bf q})] + I_{\rm bd}[m_{\bf k}({\bf q})]
	\label{eq:transp0} 
	\end{align}
in which, as we have already pointed out, ${\bf q}=(q_x, q_y=n q_0)$ and $|{\bf q}|\ll k_{\sf F}$.

\begin{figure}
\begin{center}
		\includegraphics[width=0.3\textwidth]{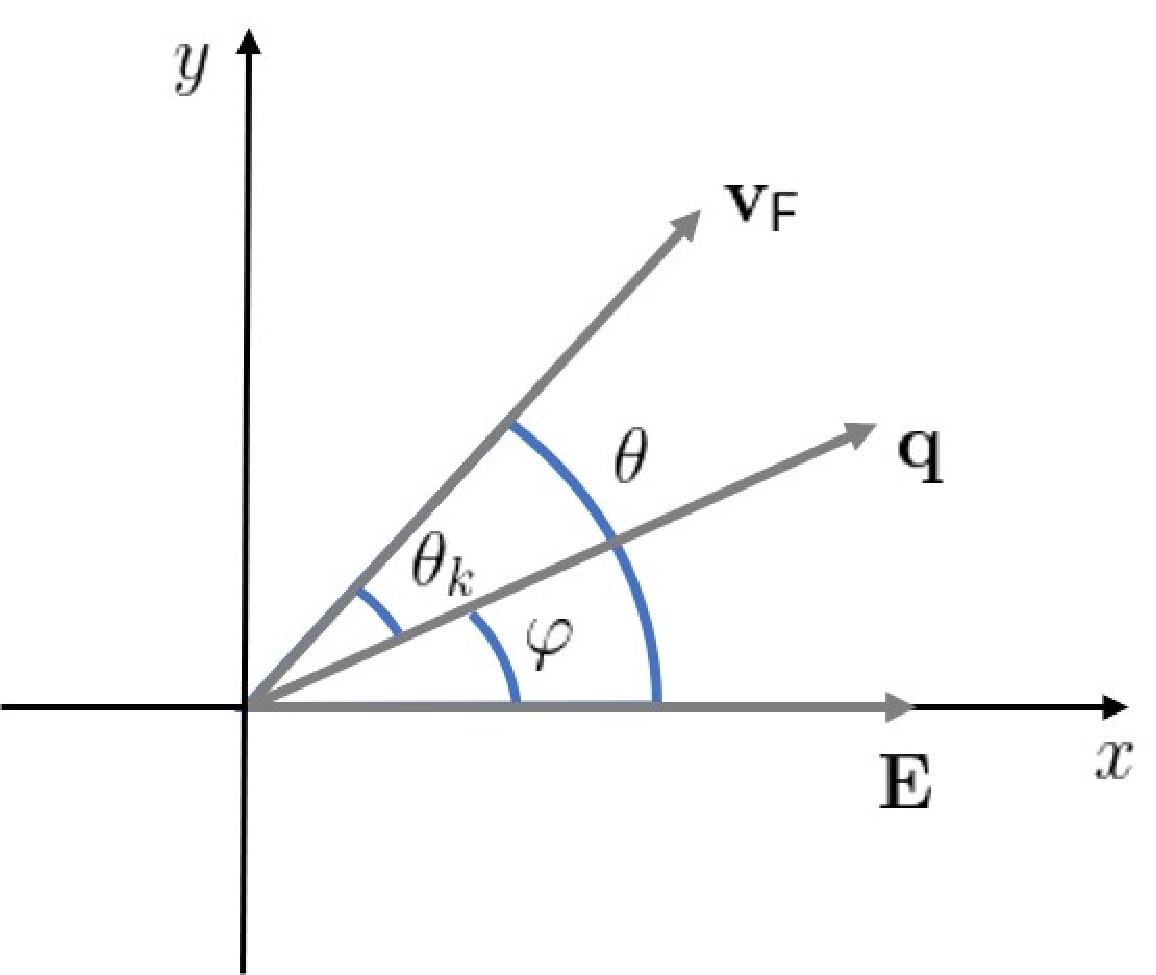}
	\end{center}
\caption{Notation convention for angles.  ${\bf E}$ is chosen in the $x$ direction.   ${\bf q}= q(\cos\varphi,\sin\varphi)$,  ${\bf v}_{\sf F}=v_{\sf F}(\cos\theta,\sin\theta)$  and  ${\bf v}_{\sf F}\cdot {\bf q}=v_{\sf F} q \cos\theta_k$. Thus,  $\theta=\theta_k+\varphi$}
\label{fig:angles}
\end{figure}

To give a more explicit expression for Eq. (\ref{eq:transp0}), we need to specify how to measure angles. 
We sketch this in Fig.  (\ref{fig:angles}).

We parametrize   ${\bf v}_{\sf F}\cdot {\bf q}=v_{\sf F} q \cos\theta_k$, ${\bf q}= q(\cos\varphi,\sin\varphi)$ and ${\bf v}_{\sf F}=v_{\sf F}(\cos\theta,\sin\theta)$, in such a way that $\theta=\theta_k+\varphi$. Thus, Eq. (\ref{eq:transp0}) takes the form, 
	\begin{align}
	&\frac{\partial\, m_{\bf k}({\bf q})}{\partial t} +i\, v_{F}\, q \cos\theta_k\left[m_{\bf k}({\bf q})+\sum_{\bf k'}F_{\bf k,k'}\,m_{\bf k'}({\bf q}) \right]
\nonumber\\
	&=e \, v_{\sf F}\, E({\bf q}) \left(\cos\varphi \cos\theta_k-\sin\varphi \sin\theta_k\right) + I_{\rm bk}[m_{\bf k}({\bf q})]  \nonumber \\ 
	&+ I_{\rm bd}[m_{\bf k}({\bf q})]
	\label{eq:transp1} 
	\end{align}
Finally, we need to specify the collision integrals. The bulk component in  the relaxation time approximation \cite{baym1991}
describes electron-impurity  collisions and it  is given by 
\begin{align}
	I_{\rm bk}[\delta n] = - \Gamma \; \delta n_{\bf k}({\bf r},t)
	\label{eq:relax}
\end{align}
where $\Gamma$ is the collision rate. Electron-electron collisions could lead the system to an electronic hydrodynamic regime where interesting features associated with collective modes could appear\cite{Krishna-2017}. However, electron-electron collision rate scales with $T^2$. Therefore, in this paper we will disregard this mechanism.
In the same approximation,  the boundary collision term is a linear function of $\delta n_k$.  It reads
\begin{align}
	I_{\rm bd}[\delta n_k({\bf r})] &=  b\,\delta\left( |y|-\frac{W}{2} \right) \nonumber 	\\
	&\times \sum_{\ell=\pm 1} \int \frac{d\theta_{k'}}{2\pi} e^{i\ell(\theta_k - \theta_{k'})}\delta n_{\bf k'}({\bf r}).
	\label{eq:bound}
\end{align}
where $\cos\theta_{k'}= {\bf k'}\cdot {\bf q}/k' q$.   The delta function has support on the boundaries $y=\pm W/2$.
The transparency factor $b$~\cite{Sodemann-2020,Guo-2017,Alekseev-2018} codifies the impurity collision rate at the boundary of the sample.   
The physical interpretation of Eq. (\ref{eq:bound}) is that impurities at the boundaries  damp linear momentum fluctuations,  given by 
$\sum_k e^{\pm i\theta_k}\delta n_{\bf k}({\bf q})$.

Fourier transforming Eqs. (\ref{eq:relax}) and  (\ref{eq:bound}),  and substituting them into Eq. (\ref{eq:transp1})
we find the explicit equation for $m_k({\bf q},t)$,  
\begin{align}
	&\frac{\partial\, m_{\bf k}({\bf q})}{\partial t} +i\, v_{F} q \cos\theta_k\left[m_{\bf k}({\bf q})+\sum_{\bf k'}F_{\bf k,k'}\,m_{\bf k'}({\bf q}) \right]=\nonumber\\
	 &-\Gamma\,m_{\bf k}({\bf q})
	+e  E({\bf q}) \left(\cos\varphi \cos\theta_k-\sin\varphi \sin\theta_k\right)- \frac{b}{W} \times \nonumber \\ 
	&\sum_{q_y',\ell=\pm 1}\int \frac{d\theta_{k'}}{2\pi} e^{i\ell(\theta_k - \theta_{k'})}\cos\Bigg[\frac{\pi(q_y-q'_y)}{q_0}\Bigg] m_{\bf k'}(q_x, q_y')
	\label{eq:transp} 
\end{align}
This is a linear equation for $m_k({\bf q},t)$,  however it is not trivial.   The interaction term $F_{{\bf k},{\bf k'}}$ mixes all points of the Fermi surface.   This is typical of Fermi liquid interactions.  On top of that,  the last term,  coming from the boundary collision integral,  is not local in $q_y'$ ,  mixing in this way all possible values of the transverse momentum. 

It is possible to give a more appealing form to this equation by re-writing it in an angular  momentum bases.  Observing that 
$m_{\bf k}({\bf q})$ is periodic in ${\bf k}$ due to the closed Fermi surface,  we can expand   
\begin{align}
	&m_{\bf k}({\bf q},t) =  m_0({\bf q},t)\ +\nonumber \\
		&\sum_{\ell=1}^\infty \left\{ m^+_{\ell}({\bf q},t)
		\cos(\ell\theta_k)+ m^-_{\ell}({\bf q},t) \sin(\ell\theta_k)\right\}. 
		\label{eq:m-Fourier} 
	\end{align}
The symmetric modes $m^+_{\ell}=m^+_{-\ell}$ parametrize the longitudinal polarization channel and the antisymmetric modes $m^-_{-\ell}=-m^-_{\ell}$ parametrize the transverse one.
Using the same reasoning, we note that $F_{{\bf k},{\bf k'}}=F(\theta_k-\theta_{k'})$ and it is periodic in these variables. Thus, we can expand it as
\begin{equation}
F_{\bf k, k'}= F_0 + \sum_{\ell=1}^{\infty}F_l\,\cos[\ell(\theta_k-\theta_{k'})]
 \label{eq:F-Fourier}
\end{equation}
where $F_\ell = F_{-\ell}$, with $\ell=0, 1,2, \ldots$, are the well known dimensionless Landau parameters of a Fermi liquid.

Rewriting Eq.~(\ref{eq:transp}) using the multipole expansions given by Eqs.  (\ref{eq:m-Fourier}) and  (\ref{eq:F-Fourier}), we find
\begin{widetext}
\begin{align}
&	\frac{\partial^2 m^\pm _\ell({\bf q},t)}{\partial t^2} + 2\Gamma\,\frac{\partial m_{\ell}^{\pm }({\bf q},t)}{\partial t} +  \Omega_{\ell}^2\, m^\pm _\ell({\bf q},t) + C_{\ell-1} m^\pm _{\ell-2}({\bf q},t)  +C_{\ell+1}\;m^\pm _{\ell+2}({\bf q},t) =\big(\partial_t + \Gamma\big) K_{1}^\pm ({\bf q},t)\delta_{1,\ell}, 
\label{eq:osci} 
\end{align}
\end{widetext}
where $\delta_{1,\ell}$ is the Kronecker delta, and,
\begin{align}
	\Omega_{\ell}^2 &= \left(\frac{v_{\sf F} q}{2}\right)^{2}\alpha_\ell(\alpha_{\ell-1}+\alpha_{\ell+1}) + {\Gamma}^2
	 \label{eq:omega} 
	 \end{align}
	 \begin{align}
	C_{\ell} &=\left(\frac{v_{\sf F} q}{2}\right)^{2}\alpha_{\ell}\sqrt{\alpha_{\ell+1}\alpha_{\ell-1}}
	\label{eq:coupling} 
\end{align}
\begin{align}
	K_{1}^+({\bf q},t) &= \frac{1}{2}\, e\,E_x({\bf q},t)\cos\varphi \nonumber \\ &- \frac{b}{W}\sum_{q_y'}\cos\left[\frac{\pi}{q_0}\left(q_y-q_y'\right)\right]m_{1}^+(q_x,q_y',t)
\label{eq:driven} 
\end{align}
\begin{align}
K_{1}^-({\bf q},t) &= \frac{i}{2}\, e\,E_x({\bf q},t)\sin\varphi \nonumber \\ &- \frac{b}{W}\sum_{q_y'}\cos\left[\frac{\pi}{q_0}\left(q_y-q_y'\right)\right]m_{1}^-(q_x,q_y',t)
\label{eq:driven-}
\end{align}
where we have introduced the coupling constants $\alpha_\ell=1+F_\ell$.

Eq.  (\ref{eq:osci}),  with the definitions of Eqs.   from (\ref{eq:omega}) to (\ref{eq:driven-}),  represents a system of damped driven coupled harmonic oscillators.  The natural frequency  $\Omega_\ell$ of each mode $m^\pm_\ell$   is given by Eq. (\ref{eq:omega}) in terms of the coupling constants $\alpha_\ell$.  Moreover,  the coupling between modes are given by Eq. (\ref{eq:coupling}).  
The damping is described by two terms:  the second term of Eq. (\ref{eq:osci}),  proportional to $\Gamma$,  codifies the effect of  bulk impurities.  On the other hand,  the second term of $K_1^\pm$ (Eqs. (\ref{eq:driven}) and (\ref{eq:driven-})),  proportional to $b/W$, includes the boundary  collision effects.

Each mode $m_\ell^\pm({\bf q},  t)$ parametrizes a Fermi surface deformation with a definite symmetry.   $m^+_\ell$  describe symmetric deformation in relation to the direction of the particle-hole momentum ${\bf q}$,  while $m^-_\ell$ represent anti-symmetric deformations related to the same axes.  On the other hand,  the index $\ell=0,  1,  2, \ldots$,   parametrizes different components of the angular momentum excitations,  which are reflected in the residual symmetries of the Fermi surface deformation.  For instance,   $m_0$,  represents isotropic deformations (or charge density excitations); it is invariant under continuous rotations.     $m^\pm_1$ represents  dipolar excitations; it has  vector character and,  as a consequence,  it is  invariant under 
$2\pi$ rotation.   Moreover,  $m_2^\pm$ denotes quadrupolar collective excitations.   Symmetric and antisymmetric components built up  a traceless symmetric second rank tensor and then it is invariant under rotations by $\pi$ (nematic symmetry).     

Some observations are in order.  Due to parity symmetry,  symmetric  ($ m^+_\ell$)  and antisymmetric ($ m^-_\ell$) modes,  as well as odd and even angular momentum modes,  are completely decoupled.  This structure enormously simplifies the analysis of the dynamics. The presence of a magnetic field breaks parity,  mixing all modes in a nontrivial way~\cite{BaFrRi2018}. On the other hand,  the electric field only couples with $m^\pm_1$.  The same happens with the boundary collision term.  Moreover,  these are the only terms in Eq.  (\ref{eq:osci}) that break rotation invariance.  For this reason,  we have grouped both terms in the definition of $K_1^\pm$.  This will simplify the evaluation of Green functions.   Interestingly,  the dipolar mode $m^\pm_1$ is the only one that contributes to the current density.  In fact,  by replacing Eqs.  (\ref{eq:normal}),  (\ref{eq:m-Fourier}) and (\ref{eq:F-Fourier}) into Eq.  (\ref{eq:J}) and integrating over $k$, we find 
\begin{align}
	J_x({\bf q}) &=\frac{e \tilde n}{m} \left\{m^+_1({\bf q}) \cos\varphi+m^-_1({\bf q}) \sin\varphi \right\}
	\label{eq:Jx}  \\
		J_y({\bf q}) &=\frac{e \tilde n}{m} \left\{m^+_1({\bf q}) \sin\varphi-m^-_1({\bf q}) \cos\varphi \right\}
	\label{eq:Jy} 
\end{align}
where $m$ is the mass of the bare electron and $\tilde n=\frac{k_{\sf F}^2}{4\pi}$ is the electron density of a circular Fermi surface.   For an arbitrary orientation of the particle-hole momentum ${\bf q}$,  both components,  symmetric as well as antisymmetric,  contribute to the current density.   If we chose to apply the electric field in the $x$ direction and there is no parity or time-reversal symmetry breaking,  then $J_y({\bf q})=0$.  This condition implies that the current density $J_x({\bf q})$ is completely determined by $m_1^+$ and is given by 
\begin{align}
	J_x({\bf q}) =\left(\frac{e \tilde n}{m}\right) m^+_1({\bf q}) \sec\varphi
	\label{eq:Jxx}
\end{align}

\section{Dipole and quadrupole interactions}
\label{Sec:Model}
In general,  there can be an infinite number of Landau parameters to parametrize interactions.  However,  in practice,  only few Landau parameters are taken in account.  In fact,  most of the properties of a normal metal can be described by considering just $F_0$, {\em i.\ e.\ }, density-density interactions.  However, novel strongly correlated compounds call for the consideration of higher angular momentum interactions\cite{Lawler2006}.  
Usually, in Galilean invariant Fermi Liquids,  $F_1$  renormalizes the effective mass,  $m^*=(1+F_1)m$\cite{nozieres-1999}. So, systems with a relatively moderate electron mass renormalization, such as GaAs/AlGaAs quantum wells or MgZnO/ZnO heterojunctions, could have $F_1 \neq 0$. Moreover,  the quadrupolar interaction $F_2$ is related to  nematicity,   observed in several strongly correlated systems.
Therefore,  in this section we present a specific model of a  Fermi liquid with both, dipolar and quadrupolar interactions and show how to compute the current density and the conductivity. In terms of the coupling constants $\alpha_\ell$,  our model is characterized by two parameters $\alpha_1 =1 + F_1$ and $\alpha_2 = 1 + F_2$,  with $\alpha_\ell = 1$ for all $\ell \neq 0, 2$.

Our interest is to compute $m_1^+ ({\bf q}, \omega)$.  Since  odd and even modes are decoupled,  we can consider Eq. (\ref{eq:osci}) only for the odd components $\ell=1,3,5,\ldots$.   Fourier transforming in time and defining the dimensionless variable  $\widetilde{z} = s - i\Gamma/v_{\sf F} q$,  where  $s = \omega/v_{\sf F} q$ with  the frequency $\omega$,   we can rewrite   Eq. (\ref{eq:osci}) in matrix notation.  
\begin{align}
\sum_{\ell'}\left[G^{+}(\widetilde{z}^{\,2})\right]^{-1}_{\ell,\ell'} m^{+}_{\ell'}({\bf q}, \omega)
	&=-i\frac{1}{v_{\sf F} q} \widetilde{z} \, K_{1}^+({\bf q},\omega)\delta_{1,\ell} 
	\label{eq:I-M}
\end{align}
where
\begin{align}
	\label{eq:M+}
	&\left[G^{+}(\widetilde{z}^{\,2})\right]^{-1}_{\ell,\ell'} =\nonumber\\ 
	&\left(
	\begin{array}{cc|ccc} 
		\widetilde{z}^{\,2} - \frac{\alpha_1(\alpha_2 + 2)}{4} & - \frac{\alpha_2\sqrt{\alpha_1}}{4} & 0 & 0 & ... \\  
		- \frac{\alpha_2\sqrt{\alpha_1}}{4} & \widetilde{z}^{\,2} - \frac{\alpha_2+1}{4} & - \frac{1}{4} & 0 & \\ 
		\hline     
		0 & - \frac{1}{4} & \widetilde{z}^{\,2} - \frac{1}{2} & - \frac{1}{4} & \ldots\\
		0 & 0 & - \frac{1}{4} & \widetilde{z}^{\,2} - \frac{1}{2}  &  \\
		\vdots &  & \vdots &  & \ddots
	\end{array}
	\right).
\end{align}
We have indicated a block structure of Eq.~(\ref{eq:M+}) by means of auxiliary vertical and horizontal lines inside the matrix. The first quadrant is a $2\times 2$ matrix,  encoding the coupling of the modes $m_1^+$ and $m_3^+$.   The fourth quadrant, which is a block with infinite components,  encodes all the odd higher angular momentum modes,  acting as  a ``heat bath" for the modes of the first quadrant.  Both blocks are coupled by off-diagonal elements $-1/4$.

Using the Green function formalism and due to the fact that the in-homogeneous term of Eq. (\ref{eq:I-M})  is a vector with only one non-zero component, the solution can be formally written as
\begin{equation}
m^+_1 ({\bf q},\omega)=-i \frac{1}{v_{\sf F} q}\widetilde{z} \; G^+_{1,1}(\widetilde{z})K_{1}^+({\bf q}, \omega)
\label{eq:m1FormalSolution}
\end{equation} 
This equation is not trivial for two reasons.  Firstly,  in order to compute $G^{+}_{1,1}(\widetilde{z})$,  it is necessary to invert  the infinite matrix of 
Eq. (\ref{eq:M+}).  Secondly,   the inhomogeneous term $K_1^+$,  given by Eq. (\ref{eq:driven}),  contains $m_1^+$  summed over all the  possible values of the transverse momentum $q_y$.  Therefore,  this is an implicit non-local equation for $m^+_1({\bf q}, \omega)$.  In the following subsections we treat both problems separately. \\

\subsection{Computation of $m_1^+(\mathbf{q}, \omega)$} 
Using Eq. (\ref{eq:driven}), we can explicitly write Eq. (\ref{eq:m1FormalSolution}) as 
\begin{align}
	m_1^+ ({\bf q}, \omega) &= \frac{1}{2} \, e\,{\widetilde G}_{11}({\bf q},\omega)E_x({\bf q},\omega)\cos\varphi \nonumber\\&- \frac{bq_o}{2\pi}\,{\widetilde G}_{11}({\bf q},\omega)\, cos\left[\frac{\pi}{q_0}q_y\right] C(q_x,\omega). 
\label{eq:m1ofc} \\
C(q_x,\omega) &= \sum_{q_y'} cos\left[\frac{\pi}{q_0}q_y'\right]m_{1}^+(q_x, q_y',\omega),
\label{eq:cofm1}
\end{align}
where we have rescaled  the Green function as 
\begin{align}
	{\widetilde G}_{11}({\bf q},\omega) = -i \frac{1}{v_{\sf F} q}\widetilde{z} \, G^{+}_{11}({\bf q},\omega) 
\end{align}
Eqs. (\ref{eq:m1ofc}) and (\ref{eq:cofm1}) form a non-local linear system for the functions  $\{m_1^+ ({\bf q}, \omega) ,C(q_x,\omega)\}$.
We strength the fact that the non-locality is in the transverse momentum  $q_y$,  and it is produced by the scattering at the boundaries of the slab.  Solving the system self-consistently for $m_1^+$ we find, 
\begin{widetext}
\begin{align}
	m_1^+ ({\bf q},\omega) &= \frac{1}{2} e\,{\widetilde G}_{11}({\bf q},\omega)\left[ E_x({\bf q},\omega)\,\cos \varphi - \frac{bq_o}{2\pi} \cos\left[\frac{\pi}{q_0}q_y\right] \frac{\sum_{q_y'} cos\left[\frac{\pi}{q_0}q_y'\right]{\widetilde G}_{11}(q_x, q_y',\omega)E_x(q_x, q_y',\omega)\cos\varphi'}{1+\frac{bq_o}{2\pi}\sum_{q_y''}{\widetilde G}_{11}(q_x, q_y'',\omega)} \right].
	\label{eq:m1solution}
\end{align}
\end{widetext} 
where $\varphi'$ is the angle subtended by the electric field $E_x$ and the momentum $(q_x, q_y')$. 
Thus,  as expected from linear response theory,   the symmetric mode $m_1^+$ is a linear function of the applied electric field. The first term of Eq. (\ref{eq:m1solution}) is local in frequency and momentum.   The coefficient $\tilde G_{1,1}$ will be related to the bulk component of the conductivity tensor.  On the other hand,  the second term is non-local in the transverse momentum $q_y$,  encoding the effect of the boundaries on the conductivity.     If we choose to apply an  electric field in $x$ direction,  which is homogeneous in the transverse $y$ direction,  $E_x(q_x,	q_y)
\equiv E_x(q_x) \delta(q_y)$,    Eq. (\ref{eq:m1solution})  localizes,   getting the simpler expression 
\begin{align}
	&m_1^+ (q_x,  nq_0, \omega) = \frac{1}{2} e\,{\widetilde G}_{11}(q_x,\omega)  E_x(q_x,\omega)
	\nonumber   \\
	& \times\left\{ \delta_{n,0} -\frac{bq_o}{2\pi} \frac{(-1)^n {\widetilde G}_{11}(q_x, n q_0,\omega)}{1+\frac{bq_o}{2\pi}\sum_{q_y''}{\widetilde G}_{11}(q_x, q_y'',\omega)} \right\}.
	\label{eq:m1solutionEx}
\end{align}
with  $n=0, \pm 1,  \pm 2, \ldots$.
In the absence of boundaries,  the only contribution is $n=0$, producing a current density $J(x,y)\equiv J_x(x)$,  uniform in $y$.    However,  the finite width of the slab induced a nontrivial distribution of $J_x(x, y)$ in the transverse direction. 
The detailed distribution depends on interactions through the structure of the Green function $\tilde G_{1,1}$.  Provided we are not interested in observing this detailed effect,  we can define a  current linear density 
\begin{equation}
J_x^{\rm linear}(x)= \int_{-W/2}^{+W/2}  J_x(x,y)\;dy 
\end{equation}
This is equivalent to observe only the $n=0$ component of Eq. (\ref{eq:m1solutionEx}).
From Eqs.   (\ref{eq:Jx})  and  (\ref{eq:m1solutionEx}) we have for the conductivity
\begin{align}
\sigma(q_x, \omega)&=\left(\frac{e^2 \tilde n}{m}\right)
\,{\widetilde G}_{11}(q_x,\omega)
\nonumber   \\
& \times\left\{ 1 -\frac{bq_o}{2\pi} \frac{ {\widetilde G}_{11}(q_x,\omega)}{1+\frac{bq_o}{2\pi}\sum_{q_y''}{\widetilde G}_{11}(q_x, q_y'',\omega)} \right\}.
	\label{eq:sigmaqx}
\end{align}

In the homogeneous limit $q_x\to 0$,  the AC conductivity finally reads, 
\begin{align}
\sigma(\omega)&=\left(\frac{e^2 \tilde n}{m}\right)
\,{\widetilde G}_{11}(0,\omega)
\nonumber   \\
& \times\left\{ 1 -\frac{bq_o}{2\pi} \frac{ {\widetilde G}_{11}(0,\omega)}{1+\frac{bq_o}{2\pi}\sum_{q_y''}{\widetilde G}_{11}(0, q_y'',\omega)} \right\}.
	\label{eq:sigmaqx0}
\end{align}
This is one of the main results of the paper.  It express the AC conductivity in terms of the Green function 
${\widetilde G}_{11}(0,\omega)$.  While the first term in Eq. (\ref{eq:sigmaqx0}) is the bulk contribution to the conductivity,  the second term codifies the boundary effects.  We will see that  this last term is responsible for the signatures of the Fermi liquid collective mode structure.

\subsection{Green Function}
It is clear from Eq.  (\ref{eq:m1solution}) or Eq.  (\ref{eq:m1solutionEx}) that the Green function $G^+_{1,1}({\bf q}, \omega)$ plays a central role in the determination of transport properties in a Fermi liquid. 
In order to compute it,  we have to invert the infinite range matrix given by Eq.~(\ref{eq:M+}) and pick up just the first component.  
There are several methods to compute this quantity.  Here we closely follow the decimation procedure described in ref. \onlinecite{Aquino-2019}. 

The main idea of the decimation technique is actually simple.   In practice,   we truncate the matrix of Eq.  (\ref{eq:M+}) to order $n\geq 2$.  Then,  we compute the inverse by usual procedures.  Due to the band structure of the matrix,  there is a recurrence relation between the inverse of the matrix truncated to order $n+1$ and  that one truncated to order $n$.    Formally, 
\begin{equation}
	{\bf G}^{+ (n+1)}={\cal  F}\left[{\bf G}^{+ (n)}\right]
\label{eq:Gtruncated}
\end{equation} 
where ${\bf G}^{+ (n)}$ and ${\bf G}^{+ (n+1)}$  are the inverse matrices of Eq.  (\ref{eq:M+}) truncated to order $n$ and $(n+1)$, respectively.   ${\cal F}$ is a well behaved matrix function.  The calculation is completed by taking  the limit
\begin{equation}
{\bf G}^{+}=\lim_{n\to \infty} {\bf G}^{+ (n)}
\end{equation}
in Eq. (\ref{eq:Gtruncated}).  We obtain an algebraic equation for  the the exact Green function ${\bf G}^{+}$. 
\begin{equation}
	{\bf G}^{+}={\cal  F}\left[{\bf G}^{+}\right]
\label{eq:Algebriac}
\end{equation} 
From its solution we can read the first element $G^+_{1,1}$.  The result is the following: 
\begin{align}
	G_{11}^{+}(\widetilde{z}) = \frac{\displaystyle \widetilde{z}^{\,2} - \frac{(\alpha_2 + 1)}{4}- \Pi(\widetilde{z})}{D^{+}(\widetilde{z})},\label{eq:exact-GF}
\end{align}
with the denominator
\begin{align}
	D^{+}(\widetilde{z}) &= \left(\widetilde{z}^{\,2} -\frac{\alpha_1(\alpha_2+2)}{4} \right)\left(\widetilde{z}^{\,2} - \frac{(\alpha_2 + 1)}{4} - \Pi(\widetilde{z}) \right)\nonumber\\ &- \frac{\alpha_1 \alpha_2^2}{16}
\label{eq:D+}
\end{align}
and
\begin{align}
	\Pi(\widetilde{z}) = \frac{1}{2} \left\{\widetilde{z}^{\,2} - \frac{1}{2} \pm \sqrt{\widetilde{z}^{\,2} \left(\widetilde{z}^{\,2} - 1\right)} \right\}. 
\label{eq:pi}
\end{align}
Eq. (\ref{eq:pi}) shows up  a branch point singularity at $\widetilde{z}=1$.  In the absence of collisions, this point is simply 
$\omega= v_{\sf F} q$, which is the threshold for Landau damping.  The square root singularity is typical of a two dimensional system.  In three dimensions,  Landau damping is characterized by a logarithmic singularity. 
This completes all the necessary ingredients to compute the conductivity $\sigma(\omega)$ given by Eq. (\ref{eq:sigmaqx0}).

\section{Collective Mode Spectrum: Exceptional Points}
\label{Sec:Collective}
We can compute the collective modes of the system by looking for the  zeroes of the denominator of the Green function, 
$D^{+}(\widetilde{z})=0$.
From Eqs.   (\ref{eq:D+}) and (\ref{eq:pi}),  we obtain
\begin{align}
	\widetilde{z}^{\,6} &-\frac{\left(\alpha_2^2 + \alpha_1\alpha_2^2 + 2\alpha_1\alpha_2 - 4\alpha_1\right)}{4 (\alpha_2 - 1)} \widetilde{z}^{\,4} \nonumber \\
	&+ \frac{\alpha_1 \left((\alpha_1 + 4) \alpha_2^2 - 4 \alpha_1\right)}{16 (\alpha_2 - 1)} \widetilde{z}^{\,2} +\frac{\alpha_1^2 \alpha_2^2}{16 (1-\alpha_2)} = 0.
	\label{eq:pole-eq}
\end{align}
which reduces to a cubic polynomial for $\widetilde{z}^{\,2}$.  
Let us explicitly show the normal modes in two simple particular cases: pure dipolar and pure quadrupolar interactions.
\subsection{Dipolar Interaction}
\label{SubSec:Dipole}
\begin{figure}
	\begin{center}
		\subfigure[]
		{\label{fig:roots-alpha1-a}
			\includegraphics[width=0.38\textwidth]{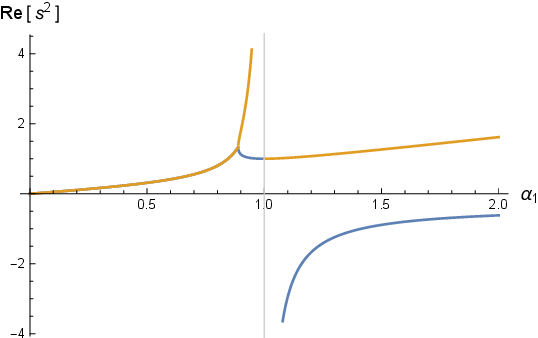}}
		\subfigure[]
		{\label{fig:roots-alpha1-b}
			\includegraphics[width=0.38\textwidth]{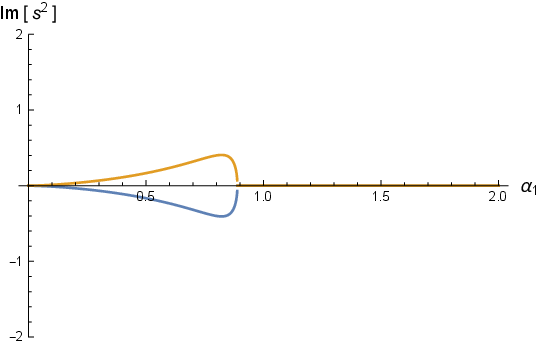}}
	\end{center}
	\caption{Symmetric collective dipolar modes. Solutions of the equation $D^+(s)=0$. 
		In panel~\ref{fig:roots-alpha1-a} we plot $\Re[s^2(\alpha_1)]$ while in panel~\ref{fig:roots-alpha1-b} we depict $\Im[s^2(\alpha_1)]$. 
		In both panels we have fixed $\alpha_2=1$ and $\Gamma=0$.}
	\label{fig:roots-alpha1}
\end{figure}

By choosing  $\alpha_2=1$,   we end up with a model with only dipolar interaction 
$\alpha_1$.  In the clean limit,   {\em i. \ e. \ }, in the collisionless regime of transport ($\Gamma = 0$),   Eq.~(\ref{eq:pole-eq}) reduces to a quadratic polynomial equation in $s^2$,  whose  roots are 
\begin{equation}
	s^2_\pm = \frac{1}{8}\frac{1}{\alpha_1-1}\left\{ 3\alpha_1^2-4\alpha_1 \pm \sqrt{9\alpha_1^4-8\alpha_1^3} \right\}.\label{eq:dipolar-poles}
\end{equation}
Interestingly,  Eq.  (\ref{eq:dipolar-poles}) displays a square-root singularity  with a branch point at $\alpha_{1}^{c} = 8/9$. This singularity is a signature of an exceptional point.  We depict the real and imaginary part of $s^2_\pm(\alpha_1)$ in Fig~(\ref{fig:roots-alpha1}). 
For repulsive interaction, $\alpha_1>1$,   the roots with $s^2>1$ is the the analog of the Landau zero sound for the dipolar interaction. The other root,   $s^2<0$ is an overdamped mode.  In the weakly attractive region,  we see two real modes.  One of them is the continuation of the Landau zero sound to the attractive region.   The other one,   with greater velocity  than the Fermi velocity can only exist in the attractive region and $s\to+\infty$ in the limit $\alpha_1\to1^-$.   For increasing values of attractive interaction,   both modes merge in the exceptional point and,  after that, they develop an imaginary part  as shown in Fig. \ref{fig:roots-alpha1}.  Therefore,  at the exceptional point,  both level crosses in the complex frequency plane.   
\subsection{Quadrupolar Interaction}
\label{SubSec:Quadrupole}
Another simple case is to consider just quadrupolar interactions.  In fact,  by solving Eq. (\ref{eq:pole-eq}) for $\alpha_1 = 1$ and  arbitrary  $\alpha_2$ we find the same qualitative structure as the dipolar case,  with the presence of an exceptional point.   In the collisioness regime and expanding the dispersion relation in the neighborhood of $s \sim 1$,   we find the roots\cite{Aquino-2020} 
\begin{equation}
s^2_\pm=\frac{1}{25}\big\{\left(16+10 \alpha_2\right)\pm \sqrt{20 \alpha_2-19}\big\}\; .
\label{eq:spm}
\end{equation} 
We see that $s^2_{\pm}(\alpha_2)$ have a square root singularity (branch point) at $\alpha_2^c=19/20$. At this point, both zeros are degenerated, $s^2_\pm(\alpha_2^c)=51/50$.
Thus, the collective modes in the pure quadrupolar case have the same qualitatively global structure than in the pure dipolar case,
 {\em i.\ e.\ }, they display two stable (real) modes for weak attraction merging into an exceptional point.
 
The exact solutions of Eq. (\ref{eq:pole-eq}) in the plane $\{\alpha_1,\alpha_2\}$ are much more involved, displaying complex  exceptional lines.  
It is worth to note that  in refs.  \onlinecite{Aquino-2019,Aquino-2020},  we have studied  a related model with charge density  and quadrupolar interactions. Interestingly,  we have found that the quadrupolar susceptibility  in the $\{\alpha_0,\alpha_2\}$ plane, also have exceptional lines displaying branch point singularities. In fact, the presence of square root singularities in the spectrum of collective modes is a general feature of two dimensional Fermi liquids with higher angular momentum interactions.

From a mathematical point of view, Eqs.~(\ref{eq:dipolar-poles}) and (\ref{eq:spm}) are simple poles of the Green's function, Eq.~(\ref{eq:exact-GF}), in two different channels: pure dipolar and pure quadrupole, respectively. At the exceptional points, $\alpha_1^c$ and $\alpha_2^c$, the Green's function have a double pole structure. Our main focus is to  study the effect of the analytic structure of the Green functions on the AC conductivity.

\section{AC  Conductivity}
\label{Sec:Conduct}
The purpose of this section is to explicitly compute the expression of  Eq. (\ref{eq:sigmaqx0}) for the AC conductivity for different interaction regimes.   We present results for  the simpler cases of pure dipolar and pure quadrupolar interactions.  The case of mixed multipolar interactions is  more involved and we will treat it in a future presentation. 

In  equation Eq. (\ref{eq:sigmaqx0}),  we can recognized two contributions. 
\begin{equation}
\sigma(\omega)=\sigma^{\rm bk}(\omega)+\sigma^{\rm bd}(\omega).
\end{equation}
The first contribution comes from the bulk and is given by 
\begin{align}
\sigma^{\rm bk}(\omega)&=\left(\frac{e^2 \tilde n}{m}\right)
\,{\widetilde G}_{11}(0,\omega).
\label{eq:sigmabk}
\end{align}
Explicitly computing the homogeneous limit of the Green function we find 
\begin{align}
\sigma^{\rm bk}(\omega)&=\left(\frac{e^2 \tilde n}{m}\right)\frac{1}{i\omega+\Gamma}
\label{eq:sigmaDrude}
\end{align}
which is the very well known Drude expression for the conductivity.   As expected,  electron-electron interactions do not contribute to the bulk conductivity.  

The second term of Eq. (\ref{eq:sigmaqx0}) comes from boundary collisions and is given by
\begin{align}
\sigma^{\rm bd}(\omega)&=-\frac{bq_o}{2\pi}\left(\frac{e^2 \tilde  n}{m}\right)  \left(\frac{1}{i\omega+\Gamma} \right)^2 \nonumber \\
&\times
\left\{ \frac{ 1}{1+\frac{bq_o}{2\pi}\sum_{q_y''}{\widetilde G}_{11}(0, q_y'',\omega)}\right\}.
\label{eq:sigmabd}
\end{align}
In order to explicitly compute this quantity we use a spectral decomposition, 
\begin{align}
\sum_{q_y''}\tilde G_{11}(0, q_y'',\omega)=\sum_j \sum_{n=-\infty}^\infty  \frac{A_j(\omega, n q_0)}{\omega-\omega_j(n q_0)-i\Gamma}
\label{eq:spectral}
\end{align}
where $\omega_j(q)$ are the collective modes represented in Fig. (\ref{fig:roots-alpha1}) and $A_j$ are the spectral weights computed by expanding  Eq.  (\ref{eq:exact-GF}) in simple fractions. 

It is convenient  to define a dimensionless expression.  For this, we rescale the conductivity as  
\begin{equation}
\sigma(\omega)=  \left(\frac{e^2 \tilde  n}{m}\right)\left(\frac{1}{v_F q_0}\right) \tilde\sigma(\omega/v_F q_0). 
\label{eq:dimensionless-Sigma}
\end{equation}
Notice that the finite size of the sample is introducing an energy scale,   $v_F q_0$,  that we are using  to scale the conductivity and the frequency,  in such a way that the dimensionless AC conductivity $\tilde\sigma$ is a function of the dimensionless frequency $\tilde\omega=\omega/v_F q_0$.  
Moreover,  we have two dimensionless parameters that characterize the collision terms:  the dissipation  in the bulk is controlled by  $\tilde \Gamma=\Gamma/v_F q_0$  and  $\gamma=b/v_F$ is related with dissipation at the boundaries of the sample. 
In this units,  the Drude conductivity reads, 
\begin{equation}
\tilde \sigma^{\rm bk}(\tilde\omega)=\frac{1}{i\tilde\omega+\tilde\Gamma}.
\label{eq:Drude}
\end{equation} 
The relation between the dimensionless parameters  $\{\tilde\Gamma, \gamma\}$  distinguishes different  electronic dynamical regimes.  When $\gamma\to 0$,  electrons rapidly scatter off of impurities and we have a diffusive transport.  In this case,  the conductivity  should follow a Drude law.   However,  when $\tilde \Gamma \approx \gamma$,  the mean free path of the quasi-particle-impurity scattering  is comparable to the material width ($W$) and  a ballistic transport regime is approached.   Moreover,  when $\tilde \Gamma \ll \gamma$, the electron–boundary scattering dominates over all others scales of the problem.  In this regime the Fermi liquid collective modes imprint very sharp signatures in the conductivity.   It is worth noting  that we have disregard quasi-particle scattering in the collision integral.  For this reason,  our results do not cover hydrodynamic transport regimes~\cite{Alekseev-2018}.

\subsection{Dipolar interactions}
\begin{figure}
	\begin{center}
\subfigure[]
		{\label{fig:sigma1P-a}		
		\includegraphics[width=0.38\textwidth]{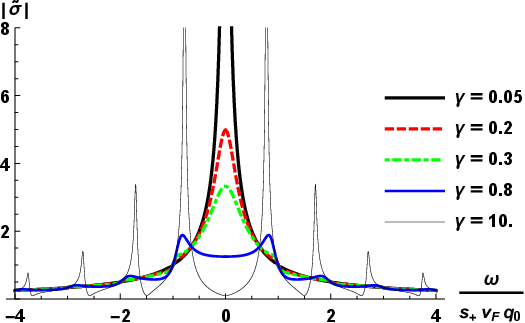}}
\subfigure[]
		{\label{fig:sigma1P-b}	
		\includegraphics[width=0.38\textwidth]{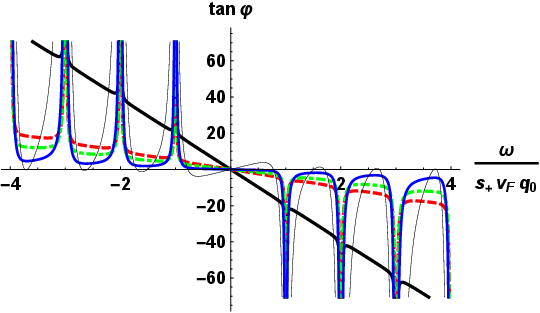}}	
	\end{center}
	\caption{Conductivity as a function of frequency in the strongly repulsive regime $\alpha_1=4$,  $\alpha_2=1$.  In this case,  the collective modes spectrum has only one stable real mode.   There are two dimensionless free parameters that controls the collision terms:  $\Gamma/v_{\sf F}q_0=0.001$ and 
	$\gamma=b/v_{\sf F}$ that we fixed as indicated in the inset of the figure.  In panel (a) we computed $|\tilde\sigma|=\sqrt{\Re(\tilde \sigma)^2+\Im(\tilde\sigma)^2}$. In panel (b) we computed the phase $\tan\varphi=\Im(\tilde\sigma)/\Re(\tilde\sigma)$. }
	\label{fig:sigma1P}
\end{figure}

Let us consider   a model with strongly repulsive dipolar interaction: $\alpha_1=4$ and $\alpha_2=1$.  As can be seen from Fig. (\ref{fig:roots-alpha1}),  there is only one stable  collective mode,  corresponding with the analogous of the Landau zero sound,  and one dissipative  mode.   In Fig. (\ref{fig:sigma1P}),  we show the dimensionless conductivity,  computed from  Eq. (\ref{eq:sigmaqx0}),  for different values of the parameters.  In panel (a) we depicted $|\tilde \sigma|=\sqrt{(\Re\tilde\sigma)^2+(\Im\tilde\sigma)^2}$,  while in panel (b),   we plotted 
 $\tan\varphi=\Im(\tilde\sigma)/\Re(\tilde\sigma)$, where $\varphi$ is the dephasing between the current density and the electric field. We have fixed $\tilde \Gamma=0.001$,  and different curves correspond to different values of the boundary parameter $\gamma=0.05,  0.2,  0.3,  0.8,10$,  ranging from a ballistic regime to a strong-boundary scattering  regime.  Moreover,  we have rescale the dimensionless frequency with the value of the pole $s_+=\pm 1.76$,   given by the positive root of Eq. (\ref{eq:dipolar-poles}).  This was done in order to simplify the visualization of novel features that are expected to appear at integer values of the rescale frequency.   

For small  values of the boundary collision term,  ($\gamma=0.05$),  the conductivity is qualitatively similar to the  Drude's law,  since the dissipation in the bulk dominates the dynamics.  For moderate values of $b/v_{\sf F}$  ($\gamma=0.2,  0.3$),  we observe a sensible reduction of the central peak  of $|\tilde \sigma|$ and very small features appear in the wings of the distribution at frequencies very near integers multiples of  the ``zero sound mode".   However,  we observe a huge signature of the presence of the collective mode in the  dephasing angle $\varphi$,  plotted in panel (b).   There is a  clear deviation from the Drude behavior,  with very well defined dips and peaks exactly at integers multiple of the resonant frequency $\omega= n s_+ v_{\sf F} q_0$,  with $n=\pm 1, \pm 2, \ldots$.   Moreover,  a very interesting  effect can be observe for greater values of $\gamma$.  For instance,  for $\gamma=0.8$,  two lateral peaks of $|\tilde \sigma|$ begin to growth,  producing a depletion at $\omega\sim 0$.  For higher multiples of the natural frequency,  other smaller peaks can be seen.  In this way,   with growing $\gamma$, the system becomes more insulating at small frequencies and the conductivity have maxima  approximately at $\omega/v_{\sf F} q_0\sim  n s_+$.  On the other hand,  the dephasing angle behaves in essentially the same way,  approximating zero for higher values of $\gamma$,  except at the integer multiples of the collective mode frequencies, where peaks and dips are developed. 
The tendency of propagating at definite frequencies increases when $\gamma$ growth.  Indeed,  for huge values of this parameter, 
for instance $\gamma=10$,  as shown by the thinner line in Fig.  (\ref{fig:sigma1P}),   the conductivity is peaked at well define frequencies.  In this regime, boundary effects dominate over the bulk dissipation,   and the transport properties strongly differs from Drude prediction.    In fact, for small frequencies,  we observe that the linear dephasing (shown in panel (b)) changes sign with respect of the Drude result.   Therefore,  in this extreme regime,  the system is almost an insulator for most frequencies,  except for some selected modes that can propagate through the strip.  In this sense,  the system turn out to be a kind of electronic wave guide.  

\begin{figure}
	\begin{center}
\subfigure[]
	{\label{fig:sigma2P-a}		
		\includegraphics[width=0.38\textwidth]{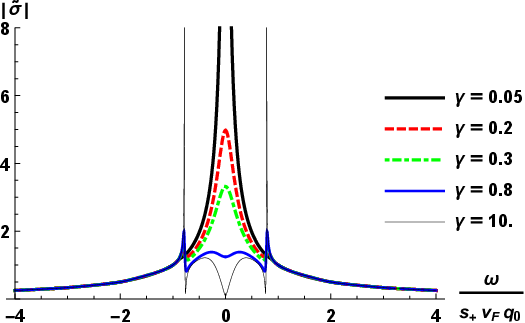}}
\subfigure[]
{\label{fig:sigma2P-b}			
	\includegraphics[width=0.38\textwidth]{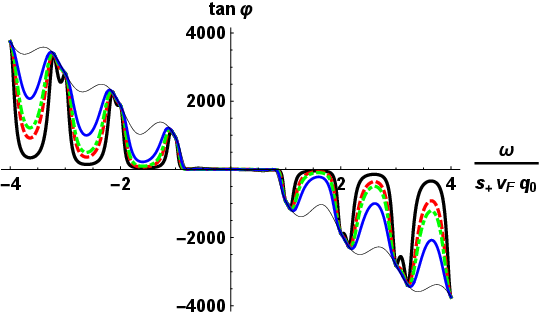}}
	\end{center}
	\caption{Conductivity as a function of frequency in the weakly attractive regime $\alpha_1=0.8899$,  $\alpha_2=1$.  In this case,  the collective modes spectrum has two stable real modes.  There are two dimensionless free parameters that controls the collision terms:  $\Gamma/v_{\sf F}q_0=0.001$ and 
	$\gamma=b/v_{\sf F}$ that we fixed as indicated in the inset of the figure.  In panel (a) we computed $|\tilde\sigma|=\sqrt{\Re(\tilde\sigma)^2+\Im(\tilde\sigma)^2}$.  In panel (b),  we computed the phase $\tan\varphi=\Im(\tilde\sigma)/\Re(\tilde\sigma)$. }
	\label{fig:sigma2P}
\end{figure}

Let us now analyze another regime.   We focus now in the weak attractive dipolar regime,  given by $\alpha_1=0.8899$ and $\alpha_2=0$.  As shown in Fig. (\ref{fig:roots-alpha1}),  $\alpha_1  \gtrsim \alpha_c=8/9$.  In this regime,   the system has two real collective modes that collapse to an exceptional point in the limit  $\alpha_1  \to \alpha_c$.  In Fig. (\ref{fig:sigma2P}) we show the conductivity using the same collision parameters than in  the previous case. 
For small and moderated values of $\gamma$,   $|\tilde\sigma|$ behaves similarly to the repulsive case.   We can observe a clear signature of the two pole structure in the dips and peaks of the dephasing angle.  However,  the valleys between dips and peaks approach zero for decreasing values of $\gamma$; this tendency is the opposite to the repulsive case. 
 For $\gamma=0.8$, a unique feature at low frequencies can be observed.  Due to the fact that the poles in the Green function are very near and the residues have opposite signs,  the conductivity develops lateral peaks whose structures in displayed in Fig.  (\ref{fig:sigma2P-a}).  For extreme higher values of $\gamma\sim 10$,  this effect saturates,   displaying two very well define peaks near  $\omega/s_+ v_{\sf F} q_0\sim \pm 1$,  and a sharp deep at $\omega=0$,  where the system become insulating. 
This unique behavior characterizes collective mode structure at the weak attractive regime. 
  
\begin{figure}
	\begin{center}
\subfigure[]
{\label{fig:sigmaEP-a}
		\includegraphics[width=0.38\textwidth]{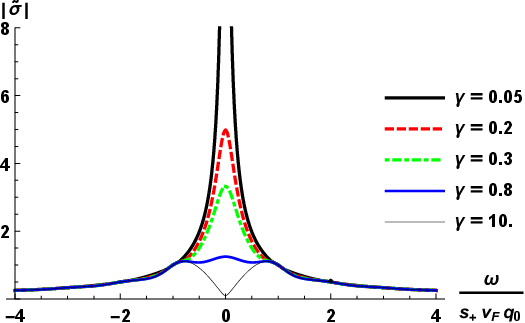}}
\subfigure[]
{\label{fig:sigmaEP-b}
		\includegraphics[width=0.38\textwidth]{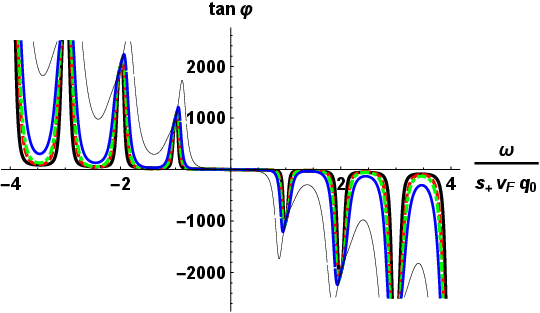}}
	\end{center}
	\caption{Conductivity as a function of frequency at the exceptional point  $\alpha_1=8/9$,  $\alpha_2=1$.   There are two dimensionless free parameters that controls the collision terms:  $\Gamma/v_{\sf F}q_0=0.001$ and 
	$\gamma=b/v_{\sf F}$ that we fixed as indicated in the inset of the figure.  In panel (a) we computed $\tilde\sigma= \sqrt{\Re(\tilde\sigma)^2+\Im(\tilde\sigma)^2}$. In  panel (b) we computed the phase $\tan\varphi=\Im(\tilde\sigma)/\Re(\tilde\sigma)$. }
	\label{fig:sigmaEP}
\end{figure}

Finally,  let us analize the exceptional point at $\alpha_1=\alpha_c=8/9$,  which is the main goal of the paper.  At this point there is a single second order pole in the Green function.  We depicted the conductivity for this case in Fig. (\ref{fig:sigmaEP}). 
The behavior of the conductivity is very similar to  the two pole strurcture at moderated values of $\gamma$.  At sufficiently strong values of boundary terms $\gamma\geq 0.8$,  $|\tilde \sigma|$ displays lateral peaks,  however much flatten than the two stable collective mode case.  Even at very strong $\gamma=10$, we observe a huge depletion at $\omega=0$, very similar to the previous case,  however with rounded lateral peaks produced by the second order pole.  The dephasing angle,  shown in Fig. (\ref{fig:sigmaEP-b}), displays the same usual dip and peak structure similar to the preceding examples,  with a different structure of the peaks.  In this case,  differently  from the simple pole of the repulsive region, the peaks are not symmetric around the collective mode frequency. 

\subsection{Quadrupolar interactions}
\begin{figure}
	\begin{center}
\subfigure[]
		{\label{fig:sigma1P-a-Q}		
		\includegraphics[width=0.38\textwidth]{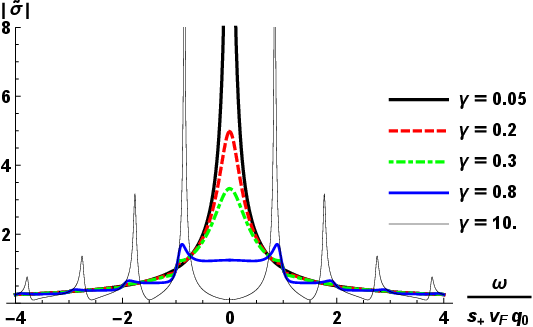}}
\subfigure[]
		{\label{fig:sigma1P-b-Q}	
		\includegraphics[width=0.38\textwidth]{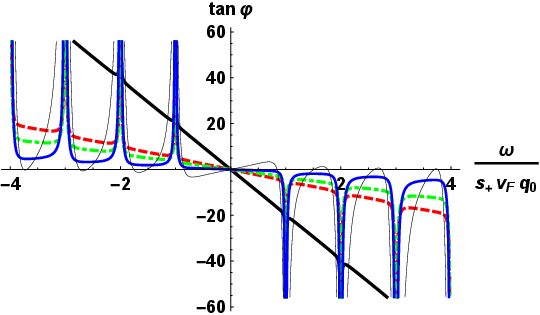}}	
	\end{center}
	\caption{Conductivity as a function of frequency in the quadrupolar strongly repulsive  regime $\alpha_1=1$,  $\alpha_2=4$.  Similarly to the dipolar case,  the collective modes spectrum has only one stable real mode.   There are two dimensionless free parameters that controls the collision terms:  $\Gamma/v_{\sf F}q_0=0.001$ and 
	$\gamma=b/v_{\sf F}$ that we fixed as indicated in the inset of the figure.  In panel (a) we computed $|\tilde\sigma|=\sqrt{\Re(\tilde \sigma)^2+\Im(\tilde\sigma)^2}$. In panel (b) we computed the phase $\tan\varphi=\Im(\tilde\sigma)/\Re(\tilde\sigma)$. }
	\label{fig:sigma1P-Q}
\end{figure}

Another simplified situation is the case of pure quadrupolar interactions $\alpha_1=1$, for arbitrary values of $\alpha_2$.
In Fig.~\ref{fig:sigma1P-Q} we show the case of strong repulsive quadrupolar interaction $\alpha_2=4$.   Since we have re-scaled the frequency in the horizontal axis with the position of the pole,  this result is essentially indistinguishable form the pure dipolar case, depicted in Fig.~\ref{fig:sigma1P}.  In fact,  it is signaling the presence of the quadrupolar equivalent of the Landau zero sound. 

\begin{figure}
	\begin{center}
\subfigure[]
	{\label{fig:sigma2P-a-Q}		
		\includegraphics[width=0.38\textwidth]{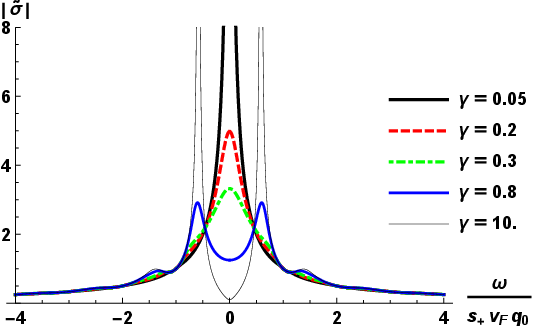}}
\subfigure[]
{\label{fig:sigma2P-b-Q}			
	\includegraphics[width=0.38\textwidth]{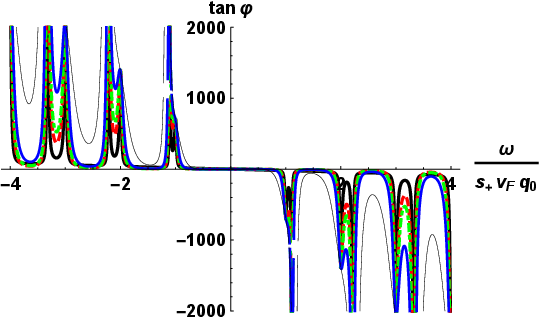}}
	\end{center}
	\caption{Conductivity as a function of frequency in the quadrupolar weakly attractive regime $\alpha_1=1$,  $\alpha_2=0.9468$.  In this case,  the collective modes spectrum has two stable real modes.  There are two dimensionless free parameters that controls the collision terms:  $\Gamma/v_{\sf F}q_0=0.001$ and 
	$\gamma=b/v_{\sf F}$ that we fixed as indicated in the inset of the figure.  In panel (a) we computed $|\tilde\sigma|=\sqrt{\Re(\tilde\sigma)^2+\Im(\tilde\sigma)^2}$.  In panel (b),  we computed the phase $\tan\varphi=\Im(\tilde\sigma)/\Re(\tilde\sigma)$. }
	\label{fig:sigma2P-Q}
\end{figure}

In Fig.~\ref{fig:sigma2P-Q},  we show the results for the weakly attractive case,   $\alpha_2=0.946819 >\alpha_c$.   
Although this figure is qualitatively similar to Fig.~\ref{fig:sigma2P},   there are some differences due to the fact that the ratio between the poles and their residues have changed.   The lateral peaks of the conductivity are more rounded than its dipolar equivalent.  However,  the dephasing angle is much more sensitive to the presence of two collective modes,  given stronger and more defined signatures. 

\begin{figure}
	\begin{center}
\subfigure[]
{\label{fig:sigmaEP-a-Q}
		\includegraphics[width=0.38\textwidth]{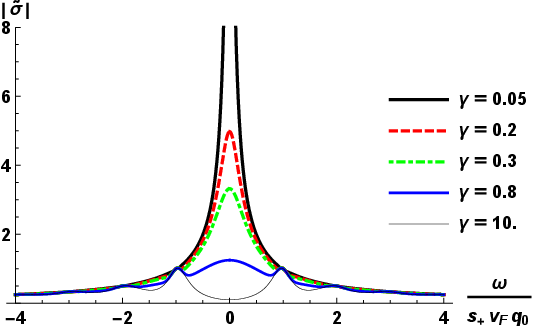}}
\subfigure[]
{\label{fig:sigmaEP-b-Q}
		\includegraphics[width=0.38\textwidth]{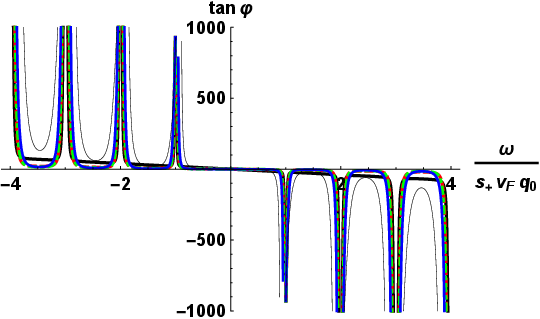}}
	\end{center}
	\caption{Conductivity as a function of frequency at the quadrupolar exceptional point  $\alpha_1=1$,  $\alpha_2=0.93681$.   There are two dimensionless free parameters that controls the collision terms:  $\Gamma/v_{\sf F}q_0=0.001$ and 
	$\gamma=b/v_{\sf F}$ that we fixed as indicated in the inset of the figure.  In panel (a) we computed $\tilde\sigma= \sqrt{\Re(\tilde\sigma)^2+\Im(\tilde\sigma)^2}$. In  panel (b) we computed the phase $\tan\varphi=\Im(\tilde\sigma)/\Re(\tilde\sigma)$. }
	\label{fig:sigmaEP-Q}
\end{figure}

Finally, in figure  \ref{fig:sigmaEP-Q},  we show the results for the quadrupolar exceptional point. 
Again, we observe the insulating character when the impurities at the border dominates the transport ($\gamma=10$),  however the peak are more rounded than the dipolar results (Fig.~\ref{fig:sigmaEP}). On the other hand,  as observed in the other interaction regimes,  the dephasing is more sensitive to the analytically structure of the Green functions, producing steepest asymmetrical peaks and deeps when the exceptional point is fined tuned.

\section{Summary and discussions}
\label{Sec:discussions}
We have computed the AC conductivity in a narrow slab in which the electronic interactions are  described by a Fermi liquid. 
Dissipation enters the Boltzmann equation through two collision integrals.  One of them represent dissipation in the bulk,  that we treated in the relaxation time approximation.  The other collision integral is concentrated at the boundaries of the slab.  

One of the main results of the paper is Eq.  (\ref{eq:sigmaqx0}), which express the AC complex conductivity in terms of the Green function of  longitudinally polarized fluctuations $G^+_{1,1}(\omega, {\bf q})$.   This expression clearly displays two contributions:  the first one has a characteristic energy scale given by the bulk dissipation $\Gamma$.  The second one comes from the boundaries of the slab.  This term is characterized by  a different energy scale,  $v_{\sf F} q_0$ where $v_{\sf F}$ is the Fermi velocity and $q_0\sim 1/W$ is fixed by the width of the slab.    

We have exactly computed the Green function $G^+_{1,1}$ for a specific  Fermi liquid model given by the the Landau parameters $F_1$ and $F_2$. Two dimensional Fermi liquids with higher angular momentum interactions display collective mode excitations whose dispersion relation has square root singularities. These complex level crossings, generally called exceptional points, appear in the weak attractive regime of definite angular momentum channels.  We have shown that, in this sense, the pure dipolar model is qualitatively similar to the pure quadrupolar one. 

Using the spectral representation of the Green function,  we computed the conductivity $|\sigma|$ and the dephasing angle $\varphi=\arctan(\Im\sigma/\Re\sigma)$. In Figs.~(\ref{fig:sigma1P}),  (\ref{fig:sigma2P}) and (\ref{fig:sigmaEP}), we show the results for different interaction regimes of a single dipole interaction model. 
 In Figs.~(\ref{fig:sigma1P-Q}),  (\ref{fig:sigma2P-Q}) and (\ref{fig:sigmaEP-Q}), we show the corresponding results for the pure quadrupolar case.  In the repulsive case,  we observe clear signatures of the ``Landau zero sound mode'' such  as well defined dips in 
$\tan\varphi$,  when the frequency  is commensurate  with the energy of the collective mode $\omega= s_+ v_{\sf F} q_0$.  On the other hand,  the conductivity has peaks for moderated and strong values of the transparency parameter $\gamma=b/v_{\sf F}$.  In the weak attractive case,   the conductivity and the dephasing clearly shows signatures of the double pole structure of the Green function.   Finally,  by fine tuning the exceptional point,  we have also a clear signature of the second order pole.   

It is worth noting that the signature of the collective mode structure in the complex AC conductivity of a Fermi liquid in clean narrow slabs is model independent.  The main result is that non-Hermitian degenerancies,  the exceptional points, which appears in a wide class of Fermi liquid models leave particular fingerprints in the conductivity,  as well as the dephasing angle.  This could be a promising tool for searching this class of Hilbert space singularities in real set ups.

We expect that  the plethora of phenomena described in this paper could be probed in different types of materials,  such as  GaAs/AlGaAs quantum wells\cite{West2010, West2011, Vakili-2004,Tan-2005,Gusev-2018},  dilute silicon\cite{Shashkin-2002,Shashkin-2003},  or MgZnO/ZnO heterojunctions\cite{Solovyev-2017,Falson_2018},  where a weak renormalization of the bare mass could indicate a fine tuning in the weak attractive dipolar region.  Moreover,  metallic compounds which presents electronic nematic phases could in principle be used to search for these signatures,  such as for instance,  Iron-based superconductors\cite{Fernandes2014,Li2017},  heavy Fermion systems such as Sr$_2$RuO$_4$\cite{Borzi2007,Wu2020} and ultracold Fermionic atoms\cite{Schunck2005,Chin2010}.

\acknowledgments
We would like to thank Inti Sodemann for bringing to our attention his paper of Ref. \onlinecite{Sodemann-2020} and for suggesting this work. 
The Brazilian agencies, {\em Funda\c c\~ao de Amparo \`a Pesquisa do Rio
de Janeiro} (FAPERJ), {\em Conselho Nacional de Desenvolvimento Cient\'\i
fico e Tecnol\'ogico} (CNPq) and {\em Coordena\c c\~ao  de Aperfei\c coamento de Pessoal de N\'\i vel Superior}  (CAPES) - Finance Code 001,  are acknowledged  for partial financial support.
RA was partially supported by a  PhD Fellowship from FAPERJ.

\end{document}